# Polarization and Anisotropy of the Microwave Sky

D. Coulson[1,2], R.G. Crittenden[1], N.G. Turok[1]

[1] *Joseph Henry Laboratory, Princeton University, Princeton NJ, 08544*

[2] *The Blackett Laboratory, Imperial College, London SW7 2BZ, UK*

(6/14/94)

## Abstract

We study the polarization-polarization and polarization-temperature correlations in standard adiabatic scenarios for structure formation. Temperature anisotropies due to gravitational potential wells and oscillations in the photon-baryon-electron fluid on the surface of last scattering are each associated with a correlated polarization pattern. While the 'correlated part' of the polarization has an r.m.s. of only a third of the total signal, it may still be measurable by mapping a large area on the sky. We calculate the expected signal to noise ratio for various measures of the polarization in a hypothetical mapping experiment such as those now being planned.

## I. INTRODUCTION

Since COBE's detection of anisotropies in the cosmic background radiation (CBR) [1], attention has focussed on obtaining higher resolution measurements, and ultimately maps of the temperature anisotropies on the sky. The angular correlations and statistics of the anisotropy pattern will yield valuable clues as to the mechanism or mechanisms involved in the formation of large scale structure in the universe - whether the perturbations were adiabatic or isocurvature, gaussian or nongaussian, whether there was a significant gravity wave component, or cosmic defects were involved. However there is substantial degeneracy amongst the predictions of different theories [2], [3], [4], and it is worth asking whether any additional information that might further discriminate between theories could be extracted from the microwave sky.

The idea that the polarization of the microwave sky might provide such additional information is not new (see e.g. [5], [6], [7], [8]), and some experimental limits have already been set [9], [10], [11]. The expected level of linear polarization is low (typically 5% of the anisotropy), but with experiments currently being planned to map the sky temperature to an accuracy of $3\mu K$ per pixel one is clearly close to the level required for a measurement. Despite the low amplitude, it has long been realized that there is a compensating factor for ground-based experiments, namely that in principle one has only to measure a single point on the sky, and in this case both polarizations are affected similarly by the atmosphere.

In this Letter, we extend previous theoretical work [7] to the temperature-polarization correlation function in standard adiabatic $\Omega = 1$ scenarios. We discuss the advantages the temperature-polarization correlation has in situations where the detector noise per pixel is larger than the signal, and show that temperature-polarization cross-correlation may be measurable in experiments with large sky coverage such as are currently being proposed.

Why should the polarization of the sky be correlated with the temperature anisotropy? At redshifts $> 1300$, the photons, baryons and electrons constitute a tightly coupled fluid. Acoustic oscillations of this fluid produced as perturbation modes cross the horizon lead to the 'intrinsic' and 'Doppler' temperature fluctuations which dominate the CBR anisotropies on small angular scales. The linear polarization produced in radiation which Thomson scatters off an electron is proportional to the quadrupole moment of the incident photon phase space density. This is vanishingly small in the fluid limit. But as photons decouple, regions in which there is a converging or diverging velocity field develop a quadrupole of order $\tau \dot{\delta}_r \sim -\tau \nabla v_r$, with $\tau$ the mean free time, $\delta_r$ and $v_r$ the density and velocity perturbation. A positive fluctuation in the radiation density $\delta_r$ produces a corresponding fluctuation in the temperature on the sky (hotter regions decouple later, and photons from them redshift less). Depending upon whether $\dot\delta_r$ is positive or negative, such a region may have a converging or a diverging velocity field. Around a temperature hot spot one would find a radial polarization pattern if $\dot\delta_r$ were positive, tangential if it were negative. Note that all modes of a given wavelength oscillate in phase, so different signs of the temperature-polarization cross correlation are expected on different angular scales.

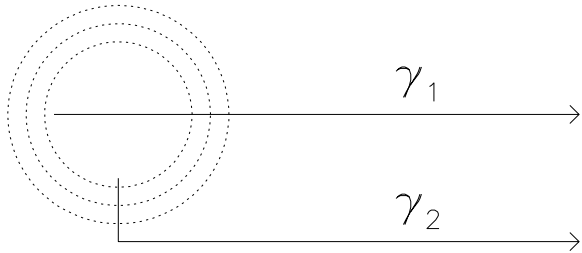

FIG. 1. A gravitational potential hill on the surface of last scattering leads to a hot spot on the sky since photons ($\gamma_1$, $\gamma_2$) blueshift as they leave it. Blueshifted photons scattered to us from one side, such as $\gamma_2$, would be linearly polarized into the page, creating a tangential pattern about the hot spot.

On larger angular scales, CBR anisotropies due to gravitational potential fluctuations on the surface of last scattering dominate in the simplest adiabatic scenarios. These also produce a correlated polarization signal (Figure 1), a tangential pattern of polarization around temperature hot spots, radial around cold spots. The scale of the polarization pattern is limited by the photon mean free path at last scattering, of order the horizon at that time - which subtends an angle of $2^o$ with standard recombination, and $6^o$ for a fully ionized universe.

## II. CALCULATION

To compute the two point correlation functions, we evolve the photon distribution function $\mathbf{f}(\mathbf{x}, \mathbf{p}, t)$ [12]. We assume the initial perturbations are adiabatic, pure growing modes, with the matter and radiation being described by perfect fluids. Here we shall present results for an $\Omega = 1$, $\Omega_B = 0.05$, $h = 0.5$ CDM dominated universe with standard thermal history, and a universe with no recombination (an extreme example of a reionized universe). Generalizations to other cosmological parameters are straightforward and will be presented elsewhere [13].

The initially Planckian, unpolarized photon distribution function is evolved forward using the general relativistic Boltzmann equation for radiative transfer with a Thomson scattering source term [14]. For simplicity we consider only scalar perturbations. In this case one needs to evolve only two transfer equations, for the components of $\mathbf{f}$ corresponding to Stokes' parameters $I$ and $Q$ [15]. These equations are written in terms of brightness functions, [16] $\Delta_i \equiv 4\delta f_i / \left(T_0 \frac{\partial \bar{f}}{\partial T_0}\right)$, where $T_0$ is the mean CMB temperature, $\bar{f}$ is the unperturbed Planck distribution, and $i = T, P$.

The computational scheme is that of Bond and Efstathiou [7], [12]. We expand the cosmological perturbations in plane waves, and the brightness functions in spherical harmonics, converting the transfer equations to a hierarchy of ordinary differential equations.

Once evolved to the present epoch, the Legendre expansion coefficients $\Delta_l^i$ ($i = T, P$) are used to construct the temperature and polarization correlation functions

$$\langle T(\hat{\mathbf{q}})T(\mathbf{e_z})\rangle = \sum_l (2l+1) C_l^T P_l(\cos\theta) \tag{1}$$

$$\langle Q(\hat{\mathbf{q}})T(\mathbf{e_z})\rangle = \cos 2\phi \sum_{l\geq 2} (2l+1) C_l^{TP} P_l^2(\cos\theta) \tag{2}$$

$$\langle U(\hat{\mathbf{q}})T(\mathbf{e_z})\rangle = \sin 2\phi \sum_{l\geq 2} (2l+1) C_l^{TP} P_l^2(\cos\theta) \tag{3}$$

$$\langle Q(\hat{\mathbf{q}})U(\mathbf{e_z})\rangle = \sin 4\phi \sum_{l\geq 4} (2l+1) C_l^{PP} P_l^4(\cos\theta) \tag{4}$$

$$\langle Q(\hat{\mathbf{q}})Q(\mathbf{e_z})\rangle = \sum_l (2l+1) C_l^P P_l(\cos\theta) + $$
$$\cos 4\phi \sum_{l\geq 4} (2l+1) C_l^{PP} P_l^4(\cos\theta) \tag{5}$$

where $(\theta, \phi)$ are the usual spherical polar angles, $\hat{\mathbf{q}} = (\sin\theta\cos\phi, \sin\theta\sin\phi, \cos\theta)$, the axes used to define the Stokes parameters are $\mathbf{e}_x$ and $\mathbf{e}_y$, and

$$C_l^T = \int k^2 dk \, |\Delta_l^T|^2 \tag{6}$$

$$C_l^P = \int k^2 dk \, |\Delta_l^P|^2 \qquad (7)$$

$$C_l^{TP} = \frac{(l-2)!}{(l+2)!} \sum_{l'} (2l'+1) \int k^2 dk \, \Delta_{l'}^P \Delta_l^{T*} a_{ll'} \qquad (8)$$

$$C_l^{PP} = \frac{(l-4)!}{(l+4)!} \sum_{l'} (2l'+1) \int k^2 dk \, \Delta_{l'}^P \Delta_l^{P*} \tilde{a}_{ll'}. \qquad (9)$$

The constants $a_{ll'}$ and $\tilde{a}_{ll'}$ are given by $a_{ll'} = \int_{-1}^{1} dx P_l(x) P_{l'}^2(x)$, and $\tilde{a}_{ll'} = \int_{-1}^{1} dx P_l(x) P_{l'}^4(x)$ which have simple closed form expressions. In deriving equations (4) and (5) we further assumed $\theta \ll 1$.

Figure 2 shows the $QT$ cross correlation function for $\phi = 0$. The $\cos 2\phi$ dependence of $\langle QT \rangle$ means that it must vanish at zero angular separation in order to be single valued. Scales $\leq 2^\circ$ are subtended by physical wavelengths shorter than the horizon at recombination, where the oscillation of the baryon and radiation fluids is reflected in the structure of the two point function. At larger angular scales, the negative tail expected from the 'gravitational potential' argument above is seen.

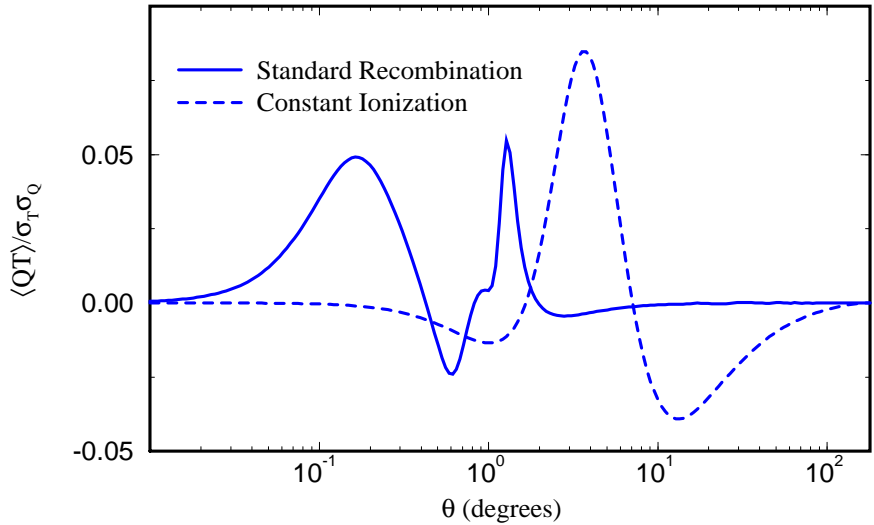

FIG. 2. $QT$ correlation function in units of $\sigma_T \sigma_Q$, where $\sigma_T^2$ is the temperature anisotropy variance and $\sigma_Q^2$ is the variance of the total polarization.

If the primordial perturbations are Gaussian, so are the temperature and polarization fields. In this case the statistics are completely described by these two point functions. We have constructed realizations of the microwave sky by making use of the small angle approximation, in which spherical harmonics may be replaced by plane waves,

$$T(\boldsymbol{\theta}) = \sum_{\boldsymbol{n}} \tilde{T}(\boldsymbol{n}) e^{i2\pi \boldsymbol{n} \cdot \boldsymbol{\theta}/\Theta} \qquad (10)$$

and similarly for $Q$ and $U$, where $\Theta$ is the angular size of the map, and the wavenumber $\boldsymbol{n}$ has integer components. Each Fourier mode $\boldsymbol{n}$ of the temperature anisotropy field $T$ is allocated a random Gaussian-distributed, complex variable $\zeta_1$ with zero mean, and unit variance $\langle \zeta_1^* \zeta_1 \rangle = 1$, according to $\tilde{T}(\boldsymbol{n}) = \sqrt{C_l^T} \zeta_1$, where $l \simeq (2\pi/n\Theta)$. A second independent random variable $\zeta_2$ is used to define the $Q$ Fourier modes:

$$\tilde{Q}(\mathbf{n}) = \left( \tilde{Q}_C(\boldsymbol{n})\zeta_1 + \tilde{Q}_U(\boldsymbol{n})\zeta_2 \right) \cos 2\phi_n \qquad (11)$$

where $\tilde{Q}_C(\boldsymbol{n})$ and $\tilde{Q}_U(\boldsymbol{n})$ are chosen to reproduce the $\langle QT \rangle$ and $\langle QQ \rangle$ correlation functions. The $U$ modes are allocated in a similar way.

The polarization field is therefore a sum of two components, which are respectively correlated, $Q_C(\boldsymbol{\theta})$, and uncorrelated, $Q_U(\boldsymbol{\theta})$, with the temperature anisotropy. Fig 3 shows the correlated component overlaid on the temperature field. The length of each vector is proportional to $[Q_C^2(\boldsymbol{\theta}) + U_C^2(\boldsymbol{\theta})]^{\frac{1}{2}}$ and the orientation is given by $2\phi = \tan^{-1}(U_C/Q_C)$. There are clear correspondences between the temperature and the correlated component of the polarization. The most obvious features are seen around hot and cold spots in an otherwise uniform background, with radial and tangential polarization patterns respectively. *Note that the Figure only shows the correlated component of the polarization - the effect of including the larger uncorrelated component is to mask almost entirely all obvious correspondences with the temperature map.*

### III. INTERPRETATION

The correlated part of the polarization is a small part of the total signal. Figure 4 shows the power spectra of the temperature-correlated polarization, $\langle Q_C Q_C \rangle$, and the total polarization, $\langle QQ \rangle$. The variance of the total polarization, $\sigma_Q^2$, corresponding to the area beneath the curve, is approximately seven times that of its correlated component, $\sigma_{Q_C}^2$. (The ratio for the constantly ionized universe is similar.) From a measured map of the temperature anisotropy, it is straightforward to construct a map of the correlated part of the polarization, as we have done in Figure 2. Because the uncorrelated part of the polarization is large, the map of the correlated part is a useful predictor of the total polarization only in a statistical sense. A one to one correspondence between features in the temperature-correlated map and those in the total polarization is not expected. However, one is much more likely to find a peak in the total polarization at a peak in the correlated map than at a random point in the sky.

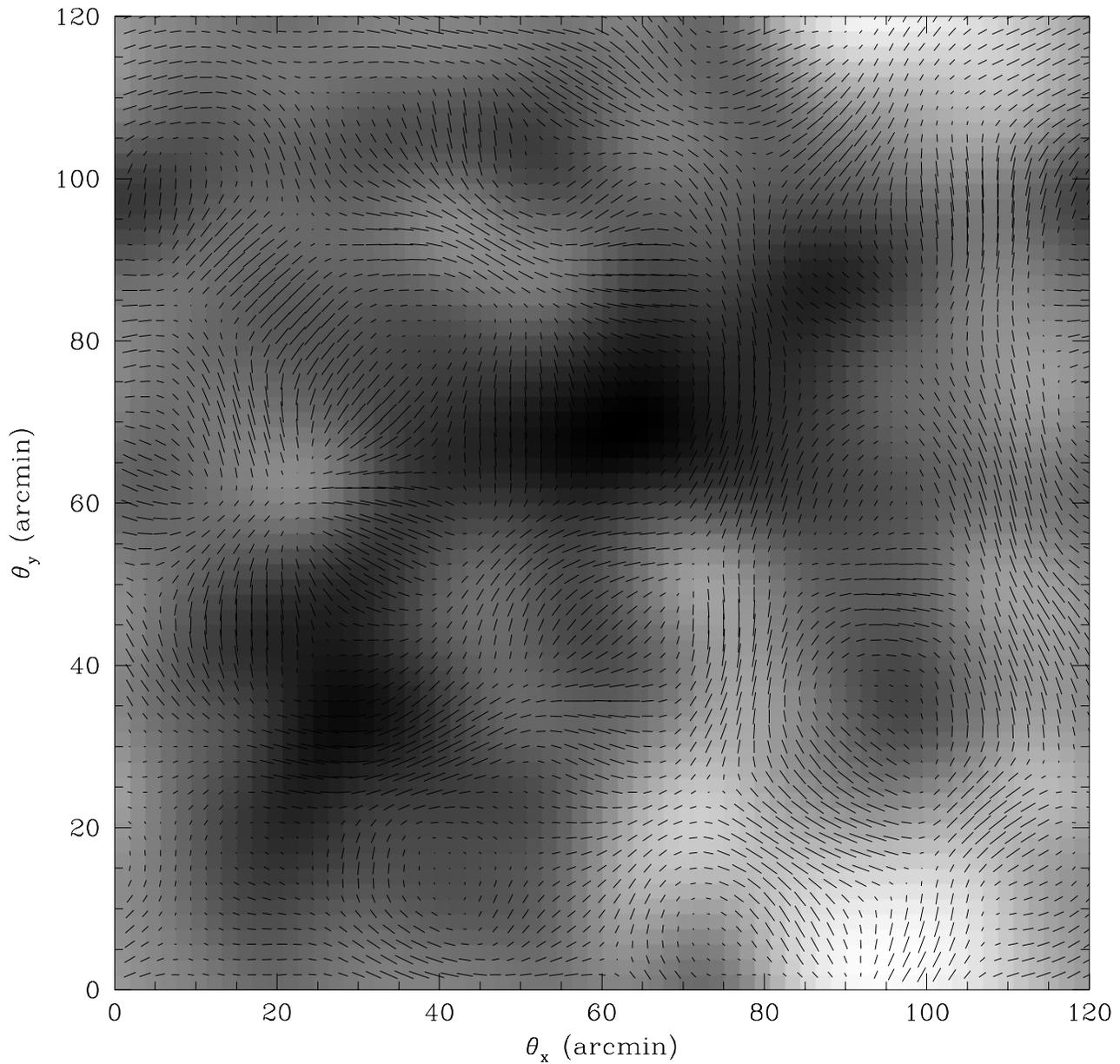

FIG. 3. $2^o \times 2^o$ temperature map (smoothed with a Gaussian beam with FWHM of $10'$) with the correlated component of the polarization overlaid. Note the cold spot at (95,35) with a surrounding tangential polarization pattern, and the hot spot at (25,65) with a radial pattern.

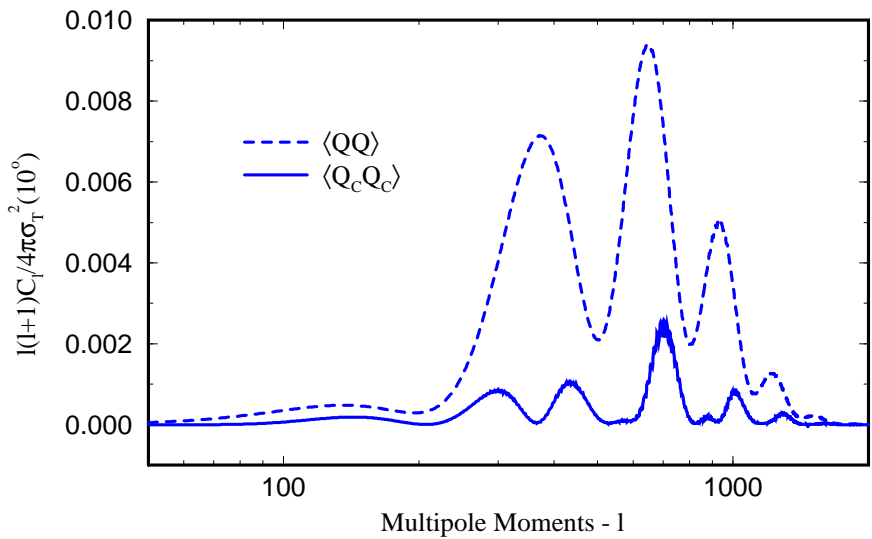

FIG. 4. The power spectra of the total and correlated polarization in a universe with a standard thermal history normalized to the COBE measurement of the temperature variance with 10° smoothing.

This statement can be quantified. The probability of finding an $n\sigma_Q$ peak in the total polarization if one is at an $m\sigma_{Q_C}$ peak in the correlated polarization is

$$p = \frac{1}{(2\pi(\sigma_Q^2 - \sigma_{Q_C}^2))^{\frac{1}{2}}} \exp[-(n\sigma_Q - m\sigma_{Q_C})^2/2(\sigma_Q^2 - \sigma_{Q_C}^2)], \tag{12}$$

while the probability of finding an $n\sigma$ peak by looking at a random point in the sky is $e^{-n^2/2}/(2\pi\sigma_Q^2)^{\frac{1}{2}}$. Thus, for example, the odds of finding a $1\sigma_Q$ peak in the total polarization at a $3\sigma_{Q_C}$ peak in the correlated polarization are three times greater than at a random point. If degree scale temperature anisotropy maps become available, it may be useful construct the map of correlated polarization (in a given theoretical scenario), and use it as a guide to the 'best' points at which to observe the polarization.

Although one expects the correlated polarization to be small, one can attempt to measure the $\langle QT \rangle$ correlation function directly. In fact, measuring $\langle QT \rangle$ could prove easier than measuring $\langle Q^2 \rangle$ because it is less susceptible to noise in the polarization measurement. Consider a detection obtained from $N$ measurements of the polarization, $Q_i \pm \sigma_D$, where $\sigma_D$ is the detector noise. If the measurements are sufficiently isolated from each other to be uncorrelated, then the measured variance will be,

$$Q^2|_{meas} = \sigma_Q^2 \pm \sqrt{\frac{2}{N}}[\sigma_Q^2 + \sigma_D^2] \tag{13}$$

where $\sigma_Q^2$ is the true polarization variance. Whereas, the measured temperature-polarization correlation will be

$$QT(\theta,\phi)|_{meas} = \langle QT(\theta,\phi) \rangle \pm \sqrt{\frac{\sigma_T^2}{N}}[\sigma_Q^2 + \sigma_D^2]^{\frac{1}{2}} \tag{14}$$

where $\sigma_T^2$ is the variance of the temperature anisotropy. (We have assumed that the noise in the temperature anisotropy is negligible.) In the limit of large detector noise, the error in measuring $Q^2$ grows as $\sigma_D^2$ while the error in $QT$ grows as $\sigma_D$, so that it becomes more advantageous to search for $QT$ correlations.

If one has a full sky temperature map but only a sparsely sampled polarization map, the noise in the temperature-polarization correlation can be reduced by including all of the temperature information. For example, we can calculate the expected correlation between the polarization at a given point and the temperature in a ring of radius $\theta$ about that point. Since the $QT$ correlation is proportional to $\cos 2\phi$, if we define $\bar{T}(\theta) = \frac{1}{2\pi} \int d\phi T(\theta,\phi) \cos 2\phi$, then $\langle Q\bar{T}(\theta) \rangle = \langle QT(\theta, \phi = 0^\circ) \rangle/2$. The error in measuring $\langle Q\bar{T}(\theta) \rangle$ is reduced significantly because $\bar{T}(\theta)$ is an average over a number of uncorrelated patches. Thus, the variance in $\bar{T}(\theta)$, $\langle \bar{T}^2(\theta) \rangle = \frac{1}{4\pi} \int d\phi \cos 2\phi C_T(\theta(2 - 2\cos\phi))$, is much smaller than the variance in the temperature anisotropy itself. For a ring of radius $\theta$, $\bar{T}$ is a weighted average of $I \sim \pi\theta/\theta_c$ independent patches (where $\theta_c$ is the temperature correlation angle) and its variance is $\langle \bar{T}^2(\theta) \rangle \sim \sigma_T^2/2I$. Assuming the rings about the polarization measurements do not have substantial overlap, this increases the signal to noise of $\langle Q\bar{T}(\theta) \rangle$ by a factor of $\sqrt{I/2}$ over that of $\langle QT(\theta) \rangle$.

Alternatively, one can test the correlated polarization map (constructed as described above) by measuring the weighted average, $\langle Q_C^{theory} Q \rangle$. For N uncorrelated measurements of the polarization,

$$Q_C^{theory} Q |_{meas} = \sigma_{Q_C}^2 \pm \sqrt{\frac{\sigma_{Q_C}^2}{N}} [\sigma_Q^2 + \sigma_D^2]^{\frac{1}{2}}. \tag{15}$$

Comparing the signal to noise of $\langle Q^2 \rangle$ and $\langle Q_C^{theory} Q \rangle$, and using $\sigma_{Q_C}^2 \sim \sigma_Q^2/7$, one finds that if the noise is greater than the polarization signal, $\sigma_D \gtrsim 1.5\sigma_Q$, then it becomes easier to measure $\langle Q_C Q \rangle$ than $\langle Q^2 \rangle$.

Consider a hypothetical experiment which measures the polarization with a 0.5° FWHM beam of 1000 well separated patches on the sky. The expected polarization for a standard recombination model is $\sigma_Q \simeq 1.4\mu K$, while the correlated polarization signal is about a third of this. If the noise level can be reduced to 3 $\mu K$ per pixel, the signal to noise of the polarization variance, $\langle Q^2 \rangle$, and that of $\langle Q_C^{theory} Q \rangle$ are both about five to one, while the signal to noise of the $\langle Q\bar{T}(\theta) \rangle$ is approximately three to one on an angular scale $\theta \simeq 1.3°$. For a fully sampled polarization map the signal to noise would improve. For the no recombination case, the expected noise levels are comparable to these because even though the variance of the polarization on 0.5° is larger ($\simeq 3\mu K$), the correlation angle is also larger and fewer independent measurements can be made from the same sky coverage. These numbers ignore any other non-cosmological sources of polarization, but on such large angular scales one can hope that astrophysical sources e.g. radio galaxies, hot gas in clusters, do not contribute significantly.

Given a fixed integration time, the optimal observation strategies for measuring the temperature polarization correlation and the polarization auto-correlation differ significantly. If the noise $\sigma_D$ is statistical, i.e. inversely proportional to the square root of the time spent in making the observation, then the best strategy for measuring the polarization variance is to spend enough time at each observation such that $\sigma_D$ is comparable to $\sigma_Q$ before moving on. The optimal strategy for measuring $\langle Q\bar{T}(\theta)\rangle$, however, is to maximize the number of observations, independent of the noise level, i.e. to map out the polarization over the whole sky. Such a strategy might most easily be realized in conjunction with a full sky anisotropy mapping experiment such as those presently being planned.

To conclude, the temperature-polarization correlation provides another observable quantity which may be used to probe the physics of the density perturbations on the surface of last scattering. Although the magnitude of the effect is small, we have shown that a reasonable signal to noise ratio is within reach of projected experiments. It would be very interesting to extend the calculations reported here to other structure formation scenarios, such as baryon isocurvature and defect models.

We thank K. Ganga, P.J.E. Peebles and D. Wilkinson for useful conversations. RC thanks R. Davis and P. Steinhardt for their collaboration in the initial development of the photon evolution code. The work of DC was supported by EPSRC (UK), while that of RC and NT was partially supported by NSF contract PHY90-21984, and the David and Lucile Packard Foundation.